\documentclass[twocolumn,aps,prb,superscriptaddress,unsortedaddress,showpacs]{revtex4}
\usepackage{graphicx}
\usepackage{epsf}
\usepackage{amsmath}
\usepackage{amssymb}

\newcommand{\etal}{{\it et al.}}
\newcommand{\sign}{\mathop{\rm sign}\nolimits}

\begin{document}

\title{Incoherent pair tunneling in the pseudogap phase of cuprates}

\author{T. Micklitz}
\affiliation{Dahlem Center for Complex Quantum Systems and Institut f\"ur Theoretische Physik,
Freie Universit\"at Berlin, 14195 Berlin, Germany}

\author{A. Levchenko}
\affiliation{Department of Physics and Astronomy, Michigan State University, East Lansing, MI 48824, USA}

\author{M. R. Norman}
\affiliation{Materials Science Division, Argonne National Laboratory, Argonne, IL 60439, USA}

\begin{abstract}
Motivated by a recent experiment by Bergeal {\it et al.}, we
reconsider incoherent pair tunneling in a cuprate junction formed
from an optimally doped superconducting lead and an underdoped
normal metallic lead. We study the impact of the pseudogap on the
pair tunneling by describing fermions in the underdoped lead with a
model self-energy that has been developed to reproduce photoemission
data. We find that the pseudogap causes an additional temperature
dependent suppression of the pair contribution to the tunneling
current. We discuss consistency with available experimental data and
propose future experimental directions.
\end{abstract}
\date{November 11, 2012}
\maketitle

\section{Introduction}

Upon lowering the temperature, the superconducting gap in underdoped
cuprates evolves smoothly from an energy gap already present in the
normal state.~\cite{timusk} Even after decades of debate, the nature
of  this `pseudogap' in the normal metallic regime of the underdoped
cuprates still remains a puzzle,~\cite{NormanPinesKallin2005} and
new experiments are needed to shed light on its nature.

One such experiment was proposed by Janko {\it et
al.}.~\cite{Boldiszar} The experimental set up consists of a
junction formed by a superconducting and a normal metallic lead in
the pseudogap phase, separated by a tunneling barrier. If the
pseudogap is due to the presence of preformed Cooper pairs, the
current-voltage ($I-V$) characteristics of such a junction should
show characteristic signatures due to pair tunneling that differ
from the standard result based on gaussian
fluctuations.~\cite{Boldiszar}

The proposed experiment was recently done by Bergeal {\it et
al.},~\cite{experiment} and their data appear to be consistent with
gaussian fluctuations. However, even if preformed pairs is not a
correct description, the pseudogap, regardless of its origin, still
affects the fermions in the normal metallic lead, and thereby the
gaussian result should not hold. A similar observation has recently
been made in the context of the Nernst effect in the pseudogap phase
of underdoped cuprates.~\cite{Nernst} Current vertices calculated
within a model that reproduces photoemission data in the pseudogap
phase~\cite{ARPES,Arc-Model} show an additional temperature
dependence, which suppresses the Nernst signal relative to the
gaussian result, consistent with experimental data.

In the present paper, we study if a similar effect of the pseudogap
changes the $I-V$ characteristics in the above mentioned tunnel
junction. We compare to experimental findings by Bergeal {\it et
al.} and discuss possible further directions to improve the
understanding of the incoherent pair tunneling in the pseudogap
phase of cuprates. Throughout the paper we set $\hbar=1$ and
$k_B=1$.

\section{Fluctuating pair tunneling}

A direct  experimental test of pairing fluctuations above $T_c$  is
the second order Josephson
effect,~\cite{Scalapino,ShenoyLee,Boldiszar,pg-JJ-noise,AL} which
has been observed in conventional superconductors \cite{goldman} and
more recently in cuprates.~\cite{experiment} The effect is exhibited
in a junction involving two leads, in the cuprate case, one
underdoped (UD), the other optimally doped (OD), with critical
temperatures such that $T_c^{UD}<T<T_c^{OD}$. The rigid pair field
of the optimally doped superconductor then plays the role of the
external field in a typical (linear response) susceptibility
measurement.

The net effect is that the fluctuating pairs  produce an additional
contribution $I_{\rm pair}$ to the current-voltage characteristics
of the junction that is directly proportional to the imaginary part
of the pair susceptibility $\chi$ of the pseudogap lead,
\begin{align}
\label{iex}
I_{\rm pair}(V,H) \propto e\, {\cal C}^2 \chi^{\prime\prime}(q(H),\omega(V)).
\end{align}
Here the frequency $\omega(V)=2eV$ is linear in the bias voltage
$V$, and momentum $q(H)$ is linear in the in-plane magnetic field
$H$. The magnitude of the pair contribution to the tunneling current
is controlled by the vertex ${\cal C}$ which describes the pair
transfer between the leads and depends on specifics of the junction.
A measurement of the excess current $I_{\rm pair}$ as a function of
$V$ and $H$ allows one to trace the frequency and momentum
dependencies of the fluctuating pair susceptibility.

\subsection{Microscopic theory}

\begin{figure}[t!]
\includegraphics[width=3in]{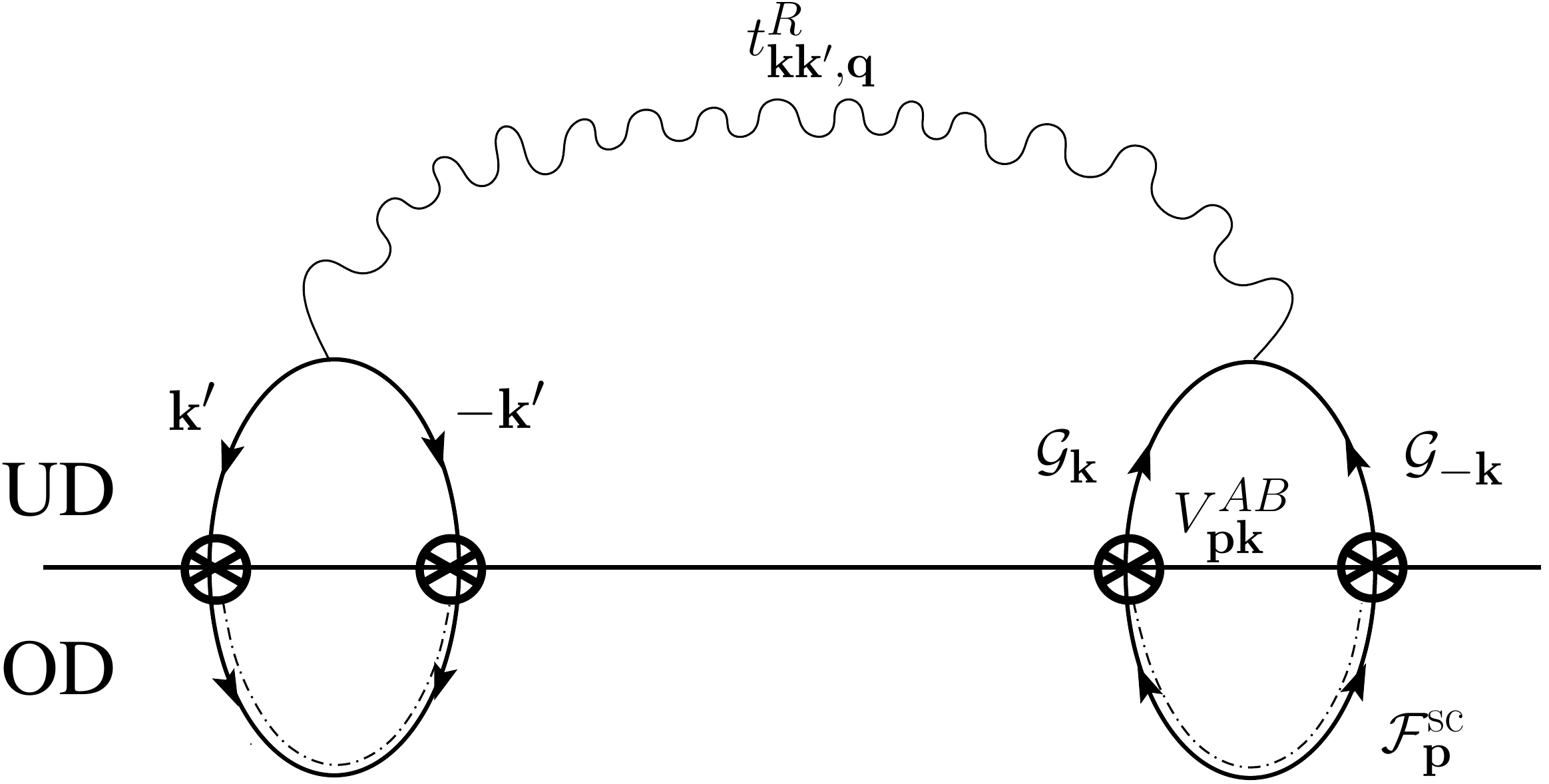}
\caption{Incoherent pair tunneling contribution,
$I_{\rm pair}=2e\, {\rm tr}\left(
F^A V^{AB} G^B G^B t^B G^B G^B V^{BA} F^A
\right)$,
to the tunneling current.
Here the trace includes summations over momenta and frequencies,
double and single lines correspond to Gor'kov functions $F^A_\bold{p}(i\epsilon_n)$ of
the optimally doped (OD) lead $A$,
and single particle Greens functions $G^B_\bold{k}(i\epsilon_n)$
of the underdoped (UD) pseudogap lead $B$, respectively,
circles represent one-electron tunneling matrix elements
$V^{AB}_{\bold{p}\bold{k}}$,~\cite{fnbarrier}
and the wavy line denotes the particle-particle
$t$-matrix for the pseudogap lead $B$ (here for q=0).
To arrive at Eq.~\eqref{excc} in the text, we
assumed a pairing instability in the $d$-wave channel
$t^R_{\bold{k},\bold{k}',\bold{q}}(\omega)=
{\cal L}_{\bold q}^R(\omega) \cos(2\varphi_\bold{k})\cos(2\varphi_{\bold{k}'})$
and kept only the relevant $d$-wave part $\propto V_1$
in the harmonic expansion of the tunneling matrix elements
$\langle |V^{AB}_{\bold{p}\bold{k}}|^2\rangle
=|V_0|^2+|V_1|^2 \cos(2\varphi_\bold{p})\cos(2\varphi_\bold{k})$, see
Refs.~\onlinecite{Boldiszar, fnbarrier}.
}
\label{fig1}
\end{figure}

In a microscopic calculation, the lowest order pair contribution to
the tunneling current arises in fourth order perturbation theory,~\cite{pt,ShenoyLee}
diagrammatically depicted in Fig.~\ref{fig1}.
Assuming a particle-particle $t$-matrix for the
pseudogap lead with pairing instability in the $d$-wave channel,
and keeping only the relevant $d$-wave part of the tunneling matrix element,
the incoherent pair contribution to the tunneling current is of the form
(see Fig.~\ref{fig1} for details)
\begin{align}
\label{excc}
I_{\rm pair}(V,H) = 4eS a^2 \, {\cal C}^2 \chi^{\prime\prime}(q(H),\omega(V))
\end{align}
where $\chi^{\prime\prime}(\bold{q},\omega)={\rm Im}{\cal
L}^R_\bold{q}(\omega)$ with ${\cal L}^R$ the retarded component of
fluctuating pair propagator, $V$ and $H$  the applied voltage and
in-plane magnetic field, and $S$ and $a^2$ the junction area and the
lattice spacing, respectively. The vertex
\begin{align}
\label{c}
{\cal C}
=
\frac{\gamma T}{N^2}\sum_{\epsilon_n}\sum_{\bold{p},\bold{k}}
& F^{\rm sc}_{\bold{p}}(i\epsilon_n,\Delta_A)
G_{\bold{k}}(i\epsilon_n) G_{-\bold{k}}(-i\epsilon_n)
\nonumber \\
&\times
\cos(2\varphi_\bold{p})
\cos^2(2\varphi_\bold{k})
\end{align}
describes the tunneling of an incoherent pair~\cite{Boldiszar} and
sets the magnitude of the pair current. We assume that the $c$-axis
is perpendicular to the junction, and all momenta and coordinates
contain only two-dimensional in-plane components. Here
\begin{align}
\label{previous}
F^{\rm sc}_\bold{p}(i\epsilon_n,\Delta) =&
{\Delta_\bold{p} \over \epsilon_n^2 +\xi_{\bold{p}}^2 +
\Delta_{\bold{p}}^2}
\end{align}
denotes the anomalous Gor'kov function of the superconducting
lead $A$, with $\epsilon_n$ (fermionic) Matsubara frequencies,
$\Delta_\bold{p}=\Delta \cos(2\varphi_\bold{p})$
with $\varphi_\bold{p}=\arctan(p_y/p_x)$,
 and $G_{\bold{k}}(i\epsilon_n)$ is the single particle propagator
of the pseudogap lead $B$ (specified below).
Finally,
$\gamma=n_i|V_1|^2/ N^2$ with $N$ the number of sites in a layer and
$n_i$ the number of impurity scattering sites per unit area of the insulating
junction.  $|V_1|$ is defined in Fig.~\ref{fig1}.~\cite{Boldiszar}

The precise form of the pair susceptibility $\chi$ varies depending
on the particular scenario adopted to describe the pseudogap phase.
In this paper, we will adopt the standard gaussian form for the pair
propagator
\begin{align}
\label{ll}
{\cal L}_{\bold q}^R(\omega )
&=
-{1\over N_0} {1\over \epsilon - i\alpha\omega + \eta q^2}
\end{align}
Here, $\epsilon=(T-T_c)/T_c$, $N_0$ the density of states,
$\alpha=\pi/8T$ and $\eta=\pi D/8T$ where $D$ is the diffusion
constant. Alternate forms, where $\alpha$ is complex (in a preformed
pairs scenario due to a BCS-BEC crossover between diffusive and
propagating pairs \cite{perali,maly,Boldiszar}) seems to be ruled
out by Bergeal \etal.~\cite{experiment}

However, even if  the pseudogap is not due to preformed pairs  and a
gaussian approach (Eq.~5) is relevant, the pseudogap will still
affect the tunneling current through the vertex ${\cal C}$. A
similar observation has recently been made in the context of the
Nernst effect in the pseudogap phase of underdoped
cuprates.~\cite{Nernst} Calculation of the current vertices within a
model~\cite{ARPES,Arc-Model} which reproduces photoemission data in
the pseudogap phase leads to an additional $T$-dependent suppression
of the Nernst effect relative to that predicted by the gaussian
model, consistent with experimental data. We will now determine if a
similar modification occurs for the tunneling current, independent
of whether the pseudogap is due to pairing or not. Before doing so,
we recall that within the $G G_0$ approximation employed by Janko
\etal,~\cite{Boldiszar} the vertex is {\begin{align} \label{c-bare}
{\cal C} &\simeq {\pi^2\over 4}n_i |V_1|^2 N_A(0) N_B(0)
\end{align}
with $N_A(0)$ and $N_B(0)$ the single-particle density of
states per spin per site for superconducting lead $A$ and pseudogap lead $B$,
respectively.

\subsection{Pseudogap vertex}

We next investigate implications of the the pseudogap for the pair
tunneling. Following Refs.~\onlinecite{Nernst, transport} we
calculate the vertex \eqref{c} using the Greens function
\begin{align}
\label{2ltm} G_{\bold{k}}(i\epsilon_n,\Delta_B) &=
{i\bar{\epsilon}_{0,n} + \xi_\bold{k} \over
(i\bar{\epsilon}_{1,n} - \xi_\bold{k})(i\bar{\epsilon}_{0,n} +
\xi_\bold{k}) - \Delta^2_{B,\bold{k}}}
\end{align}
which is based on a phenomenological self-energy describing
photoemission data in the pseudogap phase.~\cite{ARPES,Arc-Model}
Here $\Delta_{\bold k}=\Delta\cos(2\varphi_\bold{k})$ is the
momentum dependent pseudogap, $\xi_\bold{k}$ are the single particle
energies measured from the Fermi level $\mu$, and ($i=0,1$)
$\bar{\epsilon}_{i,n}=\epsilon_n + \Gamma_i \sign(\epsilon_n)$ with
$\Gamma_0$ the inverse pair lifetime proportional to $T-T_c$ (i.e.,
$\epsilon/\alpha$), and $\Gamma_1$ the single-particle scattering
rate. For $\Gamma=\Gamma_1=\Gamma_0$, Eq.~\eqref{2ltm} reduces to
the single lifetime model
\begin{align}
\label{1ltm} G_{\bold{k}}(i\epsilon_n,\Delta_B) =
-{i\bar{\epsilon}_n+\xi_\bold{k} \over \bar{\epsilon}_n^2
+\xi_{\bold{k}}^2 + \Delta_{B,\bold{k}}^2}
\end{align}
Eq.~\eqref{1ltm} gives a good description of the $T$ dependence of the Fermi arc
 if $\Gamma\propto T$.~\cite{Arc-Model}

To compute the vertex \eqref{c} we first derive that the momentum
sum for the two lifetime model is
\begin{align}
 N^{-1} \sum_\bold{k}
 &
  G_\bold{k}(i\epsilon_n)G_{-\bold{k}}(-i\epsilon_n)
\cos^2(2\varphi_{\bold{k}})
\nonumber \\
=&
{ N_B(0) \over  \Delta_B} {X_n \over 2}
\Big\{
{1\over X^2_n} \left[ E( X_n) - K(X_n) \right]
\nonumber \\
+&
\left[
1 +
Z^2_{0,n} \right] K(X_n)
-
Y^2_n Z_{1,n}Z^3_{0,n}\Pi\left(Y_n,X_n \right)
\Big\}
\nonumber \\
 \equiv \,&
 { N_B(0) \over \Delta_B}
{\cal M}^{\rm pg}(T,\Gamma_0,\Gamma_1,\Delta_B)
\end{align}
where we introduced
\begin{align}
{1\over X_n^2} &= 1 + {(\Gamma_0-\Gamma_1)^2\over 4 \Delta_{B}^2} +
{\bar{\epsilon}_{1,n} \bar{\epsilon}_{0,n} \over \Delta_{B}^2}
\\
{1\over Y^2_n} &= 1+ {\bar{\epsilon}_{1,n}\bar{\epsilon}_{0,n}
\over \Delta_B^2}
\\
Z_{i,n} &= {\bar{\epsilon}_{i,n}\over \Delta_B}
\end{align}
and $K(z)$, $E(z)$, and $\Pi(w,z)$ are the complete elliptic integrals of the first,
second and third kind.

The momentum sum over the Gor'kov Greens function, on the other hand, is
\begin{align}
 N^{-1} \sum_\bold{p}
 &
  F^{\rm sc}_\bold{p}(i\epsilon_n,\Delta_A)\cos(2\varphi_{\bold{p}})
\nonumber \\
=&
 N_A(0) k_n
\left( E(k_n) +{\epsilon_n^2\over \Delta_A^2}
\left[ E(k_n) - K(k_n) \right]
\right)
\nonumber \\
\equiv\, &
N_A(0) {\cal M}^{\rm sc}(T,\Delta_A)
\end{align}
where $k_n^2= 1/(1+ \epsilon_n^2/\Delta_A^2)$.

The vertex is then obtained by completing the Matsubara sum,
\begin{align}
\label{pgc}
{\cal C} =
&{3\pi^2\over 32} n_i |V_1|^2  N_A(0) N_B(0)
{\cal A}(T,\Gamma_0,\Gamma_1,\Delta_A,\Delta_B)
\end{align}
 Here we have included the numerical prefactor for later convenience,
 and the temperature dependent, dimensionless function
\begin{align}
\label{aex}
{\cal A}&(T,\Gamma_0,\Gamma_1,\Delta_A,\Delta_B)
\nonumber\\
&= {32 \over 3\pi^2} {T\over \Delta_B}
\sum_n {\cal M}^{\rm sc}(T,\Delta_A)
{\cal M}^{\rm pg}(T,\Gamma_0,\Gamma_1,\Delta_B)
\end{align}
It may then be verified that approximating in ${\cal A}$ the
elliptic functions by their value at zero argument describes well
the temperature dependence of Eq.~\eqref{aex} (see below). Using
that the elliptic functions at zero argument take the value $\pi/2$
we may thus approximate ${\cal
A}(T,\Gamma_0,\Gamma_1,\Delta_A,\Delta_B) \simeq
 {\cal A}_1(T/\Delta_B,\Gamma_0/\Delta_B,\Gamma_1/\Delta_B,\Delta_A/\Delta_B)$
where the temperature dependent function, normalized as ${\cal
A}_1(0,0,0,1)=1$,  ($z_n=\epsilon_n/\Delta_B$)
\begin{widetext}
\begin{align}
\label{f}
{\cal A}_1(x_0,x_1,x_2,x_3)
&=
{4 x_3 x_0 \over 3}  \sum_n \,
{
(z_n + x_1)
[ 2z_n+x_2 + x_1 ] +1
 \over
\left(
\left[
z_n^2 +x_3
\right]
\left[
(z_n+x_2)(z_n+x_1)
+ { (x_1-x_2)^2\over 4 }
+ 1)
\right]
\right)^{1/2}
\left[
(z_n+x_2)(z_n+x_1) + 1
\right]
}
 \end{align}
\end{widetext}

In the single lifetime model, we may introduce the corresponding
functions ${\cal B}(T,\Gamma,\Delta_A,\Delta_B) \equiv {\cal
A}(T,\Gamma,\Gamma,\Delta_A,\Delta_B)$ and analogously ${\cal
B}_1(T/\Delta_B,\Gamma/\Delta_B,\Delta_A/\Delta_B)$ for
Eq.~\eqref{f} in the same limit. As can be verified (see below),
taking the `zero-$T$' limit in order to convert the sum over
Matsubara frequencies into an integral gives a good description of
the vertex in the single lifetime model. We may thus further
approximate ${\cal B}(T,\Gamma,\Delta_A,\Delta_B) \simeq {\cal
B}_0(\Gamma/\Delta_B,\Delta_A/\Delta_B)$, where
  \begin{align}
\label{2ltapprox}
{\cal B}_0(x_1,x_3)
&=
{4 x_3\over 3\pi}  \int_0^\infty dz \,
{ 2(z + x_1)^2  +1
 \over
\sqrt{z^2 +x_3}
\left[ (z+x_1)^2 + 1 \right]^{3/2}
}
 \end{align}
Comparing then to  Eq.~\eqref{c-bare} in the
previous section, we find that taking into account the pseudogap
within the two and single lifetime models
\eqref{2ltm} and \eqref{1ltm} result in a
renormalization of the pair contribution to the tunneling current
by the $T$ dependent functions
${\cal A}^2(T,\Gamma_0,\Gamma_1,\Delta_A,\Delta_B)$ and
${\cal B}^2(T,\Gamma,\Delta_A,\Delta_B)$ (their approximations
${\cal A}^2_1$ and ${\cal B}^2_0$, respectively).

So far we have neglected any external magnetic fields. Variation of
an in-plane field allows one to trace the momentum dependence of the
fluctuating pair propagator, see Eq.~\eqref{excc}. If the relevant
field scale on which the pair contribution gets suppressed is small
enough so as to not affect the single particle contribution, the
magnetic field allows one to separate these two contributions to the
tunneling current. We will address this matter further in the next
section.

\begin{figure}[t!]
\includegraphics[width=3in]{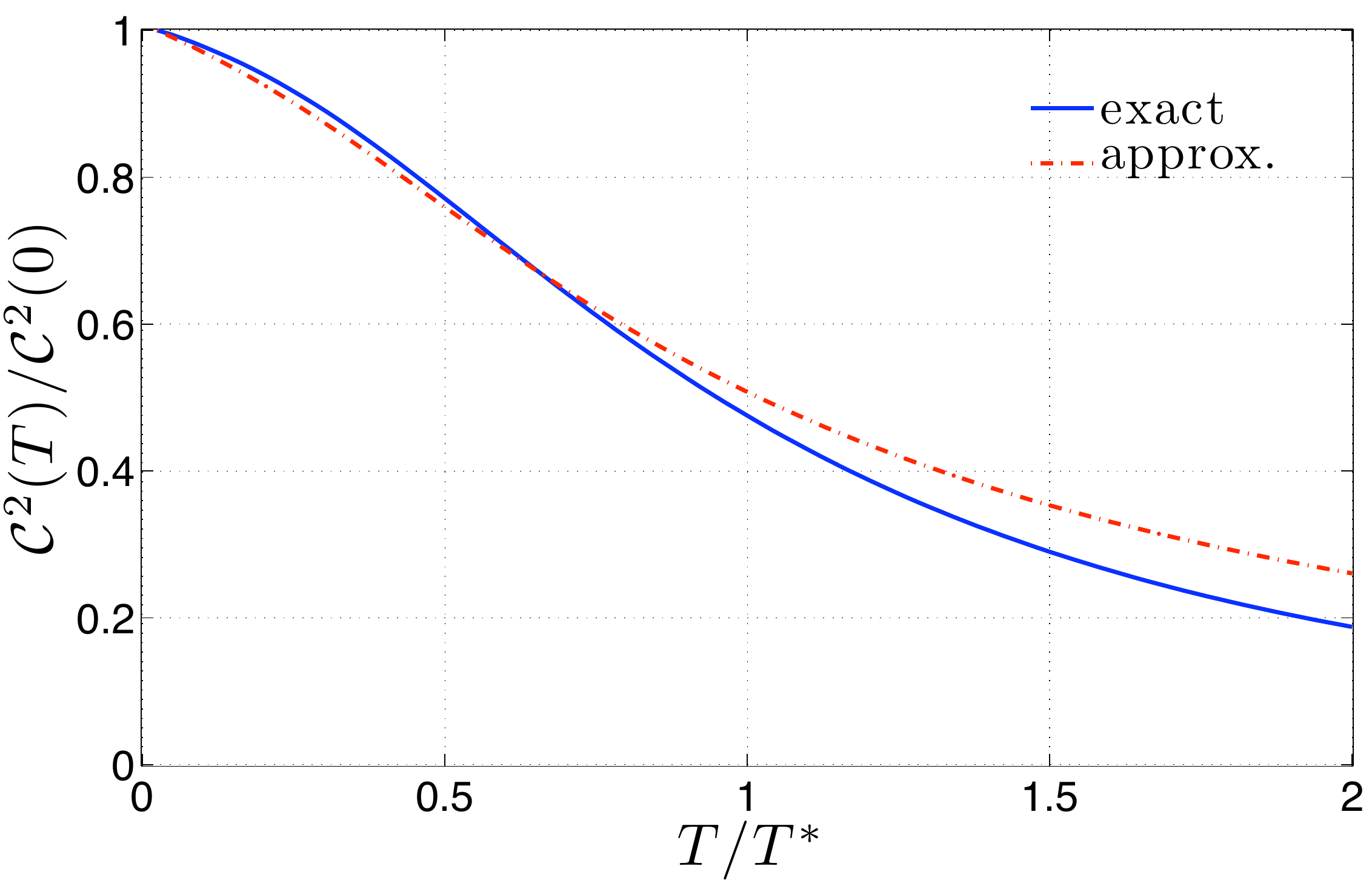}
\caption{$T$-dependence of the vertex ${\cal C}^2$ within the
single lifetime model for the pseudogap lead, where
${\cal C}\propto {\cal B}(T,\Gamma,\Delta_A,\Delta_B)$.
Here $T^*$ is the pseudogap temperature,
$\Delta_A\simeq\Delta_B=\Delta$,
 and $\Gamma/\Delta=\sqrt{3}T/T^*$
so that the Fermi arcs connect at $T=T^*$. The solid line shows the exact result
with
${\cal B}$ given in \eqref{aex} and $\Gamma_0=\Gamma_1=\Gamma$,
and the dashed-dot line shows the
approximation ${\cal B}_0$ given in \eqref{2ltapprox}.
}
\label{fig2}
\end{figure}

\subsection{Comparison to experiment}

The recent tunneling experiment by Bergeal {\it et
al.}~\cite{experiment} using a  YBCO/NdBCO junction with optimally
doped (OD) NdBCO and underdoped (UD) YBCO was designed to
test the above discussed theory. The corresponding critical
temperatures, $T^{OD}_{c}=90K$ and $T^{UD}_{c}\simeq 61K$
(the pseudogap temperature for the UD sample is $T^*\simeq250$K), allow for a
comparison to predictions in a range of temperatures extending over
a considerable fraction of $T_c$. Provided the vertex ${\cal C}$
changes only weakly with temperature, the data reported by
Bergeal {\it et al.} are consistent with a standard model of gaussian
fluctuations with susceptibility
\begin{align}
\label{l}
\chi^{\prime\prime}(q,\omega)\propto\frac{\alpha\omega}
{(\epsilon+\alpha\omega)^2+(\eta q^2)^2}
\end{align}
The extracted pair contribution to the tunneling
current is in good agreement with a Lorentzian with a width that
was 1.6 times $\Gamma_{GL}=8(T-T_c)/\pi$ at
$T-T_c=6K$ and 1.3 times at $T-T_c=9K$.
This is in good accord with ARPES, where $\Gamma_0$ was found
to be approximately twice $\Gamma_{GL}$.~\cite{ARPES}

\begin{figure}[t!]
\includegraphics[width=3in]{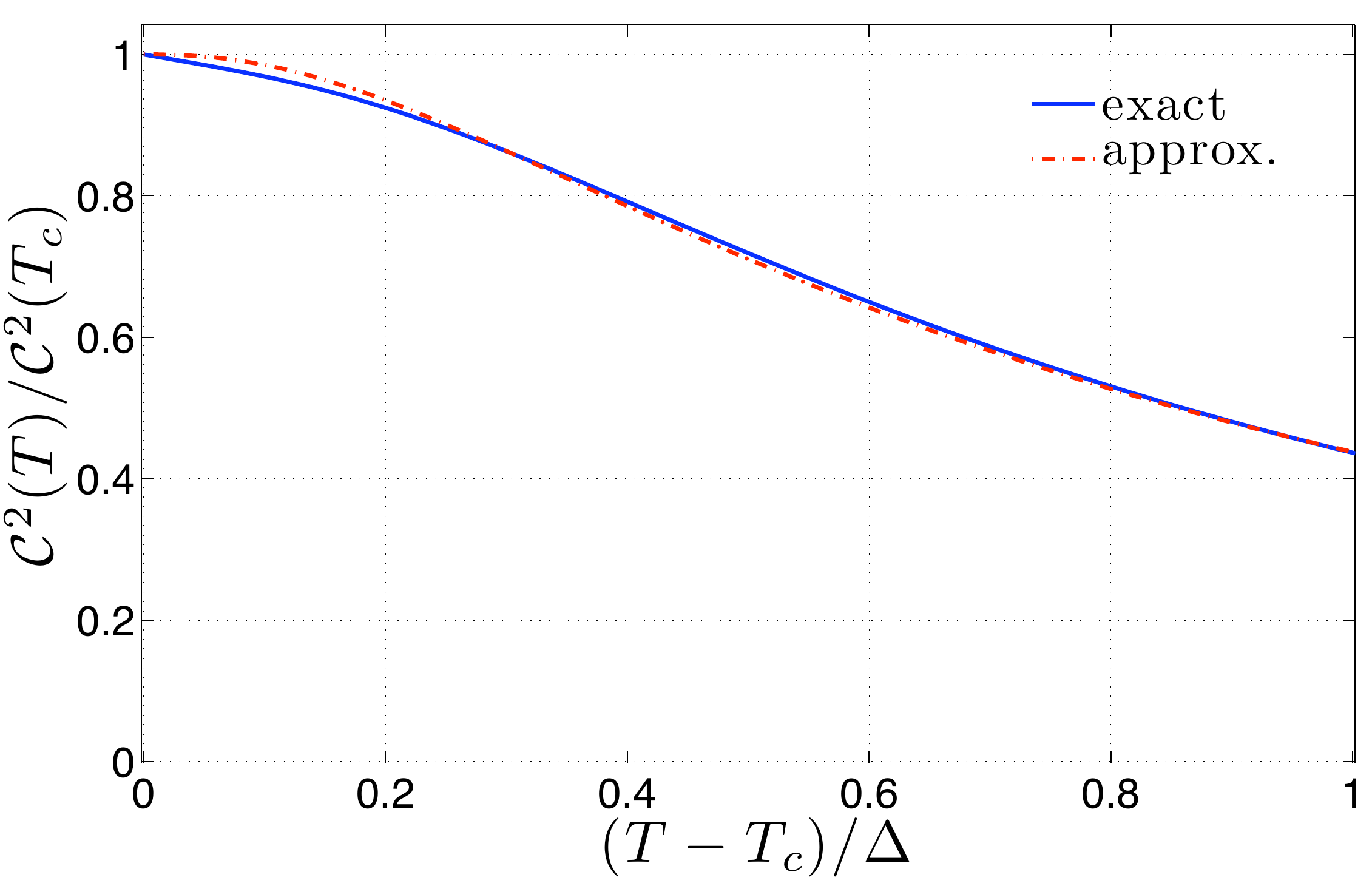}
\caption{
$T$-dependence of the vertex ${\cal C}^2$ within the
two lifetime model for the pseudogap lead, where
${\cal C}\propto {\cal A}(T,\Gamma_0,\Gamma_1,\Delta_A,\Delta_B)$.
Here $\Delta_A\simeq\Delta_B=\Delta$ and
parameters similar to Ref.~\onlinecite{ARPES},
$\Gamma_1=200 meV$, $\Delta=50 meV$,
and  $\Gamma_0=(16/\pi)(T-T_c)$, twice $\Gamma_{GL}$, where $T_c = 80K$.
 The solid line shows the exact result
  ${\cal A}$ given in \eqref{aex} and the
dashed-dot line shows the approximation ${\cal A}_1$ given in
\eqref{f}. } \label{fig3}
\end{figure}

The vertex renormalization within the single lifetime model,
${\cal B}^2(T,\Gamma,\Delta_A,\Delta_B)$,
is depicted in Fig.~\ref{fig2} for a $T$ independent
maximal value of the energy gap in the superconducting and pseudogap
 phase
$\Delta_A \simeq \Delta_B = \Delta$ with $\Gamma$ scaling as
$T$. The scaling factor used to describe the photoemission data is
$\Gamma/\Delta=\sqrt{3}T/T^*$, implying that  the $T$ dependence of
${\cal B}^2(T,\Gamma,\Delta_A,\Delta_B)$ is controlled by the
pseudogap temperature $T^*$ (at which the spectral gap `fills up' in
the antinodal region of the Brillouin zone).  For the UD sample
$T^*\simeq250$K and the temperature range in which pair tunneling is
detected by Bergeal {\it et al.} ($\sim15$K) is too narrow to
observe any noticeable deviations from predictions of the simple
gaussian formula Eq.~\eqref{l}.  To test the predicted rapid
suppression at higher temperatures coming from ${\cal
B}^2(T,\Gamma,\Delta_A,\Delta_B)$ will require using a magnetic
field to cleanly separate the pair tunneling current from the much
larger normal tunneling current.

Within the two lifetime model, the vertex renormalization depends on
the ratios $\Gamma_0/\Delta_B$ and $\Gamma_1/\Delta_B$. Since
$\Gamma_0$ has a rapid $T$ dependence near $T_c$, one might suspect
a stronger effect as compared to the single lifetime model. From
\eqref{aex} one finds, however, that for any reasonable value of
$\Gamma_1/\Delta$ comparable to that found from photoemission, the
dependence of ${\cal A}^2(T,\Gamma_0,\Gamma_1,\Delta_A,\Delta_B)$ on
temperature is qualitatively similar to that in the single lifetime
model, see Fig.~\ref{fig3}. It becomes more pronounced only for
small values of $\Gamma_1/\Delta\ll1$. We note that a zero-$T$
approximation similar to \eqref{2ltapprox} is not applicable for the
two lifetime model.

Finally, some remarks about the magnetic field dependence. The
length scale which sets the Fraunhofer pattern (i.e. the field
dependence) of the Josephson current is $Z = X_A+ X_B + d$, where
$X=\lambda\tanh(W/2\lambda)$, with $\lambda_{A,B}$ and $W_{A/B}$ the
$a/b$ penetration depth and thickness of films $A$ and $B$, and $d$
the thickness of the barrier.~\cite{weihnacht} In the two limits of
wide and thin films $X (\lambda,W)\simeq \lambda$ and  $X
(\lambda,W)\simeq W/2$, respectively. Referring then to the
experimental configuration of Bergeal {\it et al.},
 $W_A=200 nm$, $W_B=100 nm$, $d=30 nm$, and the
junction length $L=5000 nm$.  For the optimally doped
superconducting film $(A)$ $\lambda_A=100 nm$, and for the
underdoped film $(B)$ $\lambda_B=200 nm$ if it were superconducting.
However, the latter is above $T_c^{UD}$ and has no long range order,
i.e. lead $B$ is in the thin film limit $X_B\simeq W_B/2$.
Therefore, $X_A = \lambda_A \tanh(W_A/2\lambda_A)\simeq 76 nm$ and
$Z = X_A + W_B/2 + d \simeq 156 nm$. The corresponding field scale
$H$ is then estimated from $H L Z = \phi_0$ where $\phi_0$ is the
flux quantum, yielding $H\simeq 25$  Gauss. Since the normal
contribution to the current should not change much at a field of 25
Gauss, while the pair contribution is strongly suppressed at this
field, this allows one to distinguish the two contributions. This
could be exploited in future experiments.

In this context notice that, in contrast to YBCO, in the case of
BSCCO (suggested by Janko {\it et al.}~\cite{Boldiszar}), one has a
stack of Josephson junctions. The effect due to the stack would be
to create a new length scale $Z^{\prime}$ equal to the
bilayer-bilayer separation,~\cite{KleinerMuller} which is of the
order $\sim 1.5 nm$.  The resulting small length scale corresponds
to a field scale $H$ for the stack about two orders of magnitude
larger than $Z$, meaning that the presence of a stack would not
affect the $I-V$ characteristics of the A-B junction on the field
scale discussed above, and thus this complication can be ignored.

\section{Summary}

Implications of the pseudogap on transport \cite{transport} and the
Nernst effect \cite{Nernst} have been previously studied within a
phenomenological model used to describe photoemission data.  Here we
study the implications of the pseudogap on the pair tunneling, as
recently measured  by Bergeal {\it et al.}.~\cite{experiment} We
find that accounting for the pseudogap within a single lifetime
model leads to a suppression of the pair contribution to the
tunneling current relative to gaussian theory as the temperature is
increased. Within the rather small temperature range tested in the
experiment, however, this effect would not be noticeable. To
determine this would require differentiating the pair current from
the much larger normal current, which would require the application
of a small in-plane magnetic field.  We contrast this with the
Nernst effect, where the normal background is significantly smaller.
Therefore, we suggest that such field dependent measurements be done
in the future that would not only help identify effects due to the
vertex, but also test the validity of Eq.~5 in the context of
specific theories for the pseudogap phase.~\cite{she}

\acknowledgments

This work was supported by the U.~S.~Dept.~of Energy, Office of
Science, Basic Energy Sciences, under contract DE-AC02-06CH11357.
A.L. acknowledge support from Michigan State University.


\begin{thebibliography}{99}

\bibitem{timusk}
T.~Timusk and B.~Statt, Rep. Prog. Phys. {\bf 62}, 61 (1999).

\bibitem{NormanPinesKallin2005}
M.~R.~Norman, D.~Pines and C.~Kallin, Adv. Phys. {\bf 54}, 715 (2005).

\bibitem{Boldiszar}
B.~Janko, I.~Kosztin, K.~Levin, M.~R.~Norman and D.~J.~Scalapino,
Phys. Rev. Lett. {\bf 82}, 4304 (1999).

\bibitem{experiment}
N.~Bergeal, J.~Lesueur, M.~Aprili, G.~Faini, J.~P.~Contour and
B.~Leridon, Nature Phys. \textbf{4}, 608 (2008).

\bibitem{Nernst}
A.~Levchenko, M.~R.~Norman and A.~A.~Varlamov, Phys. Rev. B
\textbf{83}, 020506 (2011).

\bibitem{ARPES}
M.~R.~Norman, M.~Randeria, H.~Ding and J.~C.~Campuzano, Phys. Rev. B
{\bf 57}, R11093 (1998).

\bibitem{Arc-Model}
M.~R.~Norman, A.~Kanigel, M.~Randeria, U.~Chatterjee and
J.~C.~Campuzano, Phys. Rev. B \textbf{76}, 174501 (2007).

\bibitem{Scalapino}
D.~J.~Scalapino, Phys. Rev. Lett. {\bf 24}, 1052 (1970).

\bibitem{ShenoyLee}
S.~R.~Shenoy and P.~A.~Lee, Phys. Rev. B {\bf 10}, 2744 (1974).

\bibitem{pg-JJ-noise}
X.~Dai, T.~Xiang, T.-K.~Ng and Z.-B. Su, Phys. Rev. Lett.
\textbf{85}, 3009 (2000).

\bibitem{AL}
A.~Levchenko, Phys. Rev. B \textbf{78}, 104507 (2008).

\bibitem{goldman}
J. T. Anderson and A. M. Goldman, Phys. Rev. Lett. {\bf 25}, 743 (1970).

\bibitem{fnbarrier}
In the above experiment, it is assumed that tunneling takes place via direct
or resonant processes through localized states in the barrier, and
$\langle |V^{AB}|^2\rangle$ is the impurity averaged single electron transfer through
the diffusive tunneling barrier.

\bibitem{pt}
H.~Takayama, Progr. Theor. Phys. {\bf 46}, 1 (1971).

\bibitem{perali}
A.~Perali, P.~Pieri, G.~C.~Strinati and C.~Castellani, Phys. Rev. B {\bf 66}, 024510 (2002).

\bibitem{maly}
J.~Maly, B. Janko and K.~Levin, Physica C {\bf 321}, 113 (1999).

\bibitem{transport}
A.~Levchenko, T.~Micklitz, I.~Paul and M.~R.~Norman, Phys. Rev. B
\textbf{82}, 060502 (2010).

\bibitem{weihnacht}
M. Weihnacht,  Phys. Stat. Sol. {\bf 32}, K169 (1969).

\bibitem{KleinerMuller}
R. Kleiner and P. M\"uller, Phys. Rev. B {\bf 49}, 1327 (1994).

\bibitem{she}
J.-H. She, B. J. Overbosch, Y.-W. Sun, Y. Liu, K. E. Schalm, J. A. Mydosh and J. Zaanen,
Phys. Rev. B {\bf 84}, 144527 (2011).

\end{thebibliography}
\end{document}